\documentclass[submit]{epsv8}

\usepackage{amsmath,amssymb}
\usepackage{graphicx}
\usepackage{booktabs}
\usepackage{array}
\usepackage{float}
\usepackage{xcolor}
\usepackage{lineno}
\usepackage[hidelinks,hypertexnames=false]{hyperref}
\newcolumntype{L}[1]{>{\raggedright\arraybackslash}p{#1}}
\newcommand{\safeincludegraphics}[2][]{\IfFileExists{#2}{\includegraphics[#1]{#2}}{\fbox{\parbox{0.8\linewidth}{Missing figure file: \texttt{\detokenize{#2}}}}}}
\makeatletter
\g@addto@macro\@floatboxreset{\normalsize}
\makeatother

\title{An Approximately 70-Year Core-Related Modulation of Earth Rotation and Its Implications for the Leap Second}

\author{
Zewen Zhang$^{1,2}$, Yuanwei Wu$^{1,*}$, Xishun Li$^{1}$, Dang Yao$^{1}$, Xuan Cheng$^{1}$, \\
	Xuhai Yang$^{1}$, Shougang Zhang$^1$\\
	\footnotesize 
	$^{1}$National Time Service Center, Chinese Academy of Sciences, Xi'an 710600, China
	\\
	\footnotesize 
	$^{2}$School of Astronomy and Space Science, University of Chinese Academy of Sciences, Beijing 100049, China\\
		\footnotesize $^{*}$Correspondence: yuanwei.wu@ntsc.ac.cn
}

\abstract{Recent observations of Universal Time (UT1) indicate an acceleration in Earth's rotation. If sustained under the current leap-second framework, this behavior could eventually prompt consideration of a negative leap second. We examine whether the recent acceleration is consistent with an approximately 70-year, core-related modulation of length of day (LOD). After removal of modeled tidal, surface-fluid, and secular contributions, residual LOD contains a near-70-year component, and a similar component is present in core angular momentum (CAM)-derived equivalent LOD inferred from geomagnetic observations. All harmonic, spectral, and LOD--CAM analyses reported here use the common 1883--2022 interval. Harmonic regression over trial periods of 50--100 yr gives periods of 69.7 yr for residual LOD and 71.8 yr for CAM-derived equivalent LOD, with amplitudes of 2.87 and 1.94 ms, respectively. Lomb--Scargle spectra show peaks near 67.8 and 70.5 yr. The annual series have a zero-lag correlation of 0.918. Their lagged correlation has a broad maximum for a CAM lead of approximately 1--3 yr, with a numerical maximum of 0.932 at 2 yr. Because both records are strongly autocorrelated, these coefficients are used to characterize their correspondence rather than to assess predictive significance. The results are consistent with a core-related contribution to low-frequency rotational variability, but they do not uniquely separate the contributions of electromagnetic, topographic, gravitational, and viscous core--mantle coupling mechanisms. Within the fitted model, the multidecadal component alone does not indicate sustained near-term shortening of the day that would, by itself, require a negative leap second. This is a model-dependent geophysical assessment, not an operational prediction of future UTC adjustments.}


\begin{document}
\maketitle
%\linenumbers
\renewcommand{\arraystretch}{1.5}

	\section{Introduction}
	
	Coordinated Universal Time (UTC) is derived from atomic time and is kept close to UT1, the time scale based on Earth's rotation, through the insertion of leap seconds. These adjustments maintain $|\mathrm{UT1}-\mathrm{UTC}|$ within the prescribed tolerance, whereas International Atomic Time (TAI) remains continuous and is not adjusted to follow Earth rotation. Since the current leap-second system was introduced in 1972, 27 leap seconds have been inserted, all positive, at a mean interval of approximately 2--3 years \citep{dickIERSLEAPSECOND,nelsonLeapSecondIts2001}. The most recent adjustment occurred on January 1, 2017. 
	
\begin{figure}[H]
		\centering
		\includegraphics[width=0.52\textwidth,height=0.35\textheight,keepaspectratio]{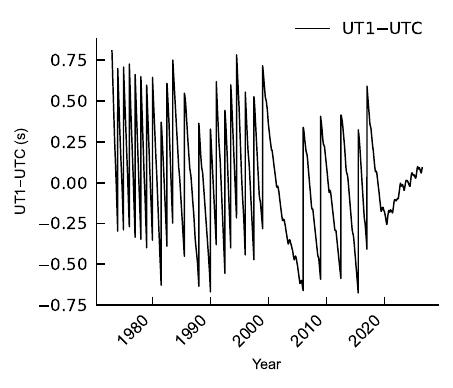}
		\caption{\textbf{UT1--UTC from 1972 through 2024.} The curve shows the difference between Earth-rotation time and UTC after leap-second adjustments. The change in slope after the early 2000s motivates examination of the low-frequency LOD components relevant to future UTC adjustments.}
		\label{fig:1}
	\end{figure}

	The slope of the UT1--UTC series (Fig.~\ref{fig:1}) changed markedly after the early 2000s and became positive around 2020. UT1 has therefore tended to advance relative to UTC, consistent with an LOD shorter than the nominal 86,400 s day and with faster rotation of the solid Earth. This recent acceleration contrasts with the long-term tidal deceleration of Earth's rotation. If it persisted, a negative leap second could eventually be considered. Such an adjustment has not previously been implemented and could create operational risks for timekeeping and satellite systems.
	
	Earth's polar motion also changed markedly near 2000, when the spin-axis trajectory shifted by approximately 75$^\circ$ toward the Greenwich meridian. Terrestrial water storage (TWS) and cryospheric mass redistribution have been estimated to account for more than 80\% of the associated amplitude and directional change \citep{adhikariClimatedrivenPolarMotion2016}. Comparable surface-mass processes, however, do not account for the recent UT1 and LOD trends \citep{zotovAnalysisEarthPolar2022}. 
	
	Over geological timescales, tidal dissipation slows Earth's rotation and increases LOD by approximately $2.3 \pm 0.1$ ms per century. Secular changes in oblateness, represented by the $J_2$ coefficient measured with Satellite Laser Ranging (SLR), also modify the axial moment of inertia, but their contribution is too small to explain most of the recent rotational acceleration \citep{mitrovicaReconcilingChangesEarth2015,yoderSecularVariationEarth1983a}. 
	
	These observations suggest an additional internal contribution to the multidecadal angular-momentum budget. Previous studies have identified 60--70 yr variability in LOD and geomagnetic quantities and have related it to core dynamics, stable stratification, or angular-momentum exchange at the core--mantle boundary (CMB) \citep{robertsOn60yearSignal2007,buffettGeomagneticFluctuationsReveal2014}. Intradecadal LOD variations have also been associated with geomagnetic jerks and core-surface flow \citep{duanIntradecadalVariationsLength2020}, while recent UT1 projections indicate that an imminent negative leap second is unlikely \citep{malkinShouldExpectFurther2024}. Here we examine whether angular-momentum exchange with the fluid outer core contributes substantially to low-frequency LOD variability. We remove the modeled tidal, surface-fluid, and secular components and compare the resulting residual with CAM inferred from geomagnetic-field inversions. The analysis is intended to constrain a core-related contribution, not to provide an operational UT1 forecast or to partition the signal uniquely among individual CMB coupling mechanisms.

	Building on these studies, we quantitatively compare residual LOD with CAM-derived equivalent LOD over their common annual interval. We estimate their periods, amplitudes, and phases using harmonic regression, Lomb--Scargle spectra, and BIC-based period selection, and then integrate the fitted LOD component to evaluate its contribution to continuous UT1. This approach links multidecadal core-related angular-momentum variability to UT1 without assigning the signal to a single electromagnetic, topographic, gravitational, or viscous coupling mechanism.

	\section{Physical Framework}

	\subsection{Axial Angular-Momentum Balance}

	Small changes in Earth's rotation are governed by the Liouville equation. In an Earth-fixed frame, the total angular momentum can be written as
	\begin{equation}
		\mathbf{H}=(\mathbf{I}_0+\Delta\mathbf{I})\boldsymbol{\omega}+\mathbf{h},
	\end{equation}
	where $\mathbf{I}_0$ is the reference inertia tensor, $\Delta\mathbf{I}$ is its perturbation, $\boldsymbol{\omega}$ is the rotation vector, and $\mathbf{h}$ is the relative angular momentum of the moving fluids. After the modeled external tidal torques are removed, axial changes in the rotation of the solid Earth are balanced by angular-momentum exchange with the atmosphere, oceans, terrestrial hydrosphere, and core \citep{gross309EarthRotation2007,barnesAtmosphericAngularMomentum1983,petitIERSConventions2010}. In anomaly form, this balance may be summarized as
	\begin{equation}
		H'_s+H'_a+H'_o+H'_{oc}+H'_{ic}=H'_e,
	\end{equation}
	where the subscripts denote the solid Earth ($s$), atmosphere ($a$), ocean and terrestrial hydrosphere ($o$), outer core ($oc$), inner core ($ic$), and the total Earth system ($e$). LOD records the axial response of the solid Earth to these exchanges.

	\subsection{Surface-Fluid and Core Contributions}

	Atmospheric winds and pressure, ocean circulation, and terrestrial water storage redistribute mass and relative angular momentum and therefore excite Earth-rotation variations \citep{gross309EarthRotation2007,deeERAInterimReanalysisConfiguration2011,jungclausCharacteristicsOceanSimulations2013,dillHydrologicalModelLSDM,hagemannDocumentationHydrologicalDischarge1998}. Their principal contributions span subdaily to interannual timescales, although lower-frequency components may remain. At longer timescales, electromagnetic, topographic, gravitational, and viscous stresses at the core--mantle boundary can transfer angular momentum between the mantle and core \citep{buffettGravitationalOscillationsLength1996,buffettGravitationalBrakingInnercore2000,chaoGravitationalEnergyAssociated2014}. The analysis below removes the modeled tidal and available surface-fluid contributions before comparing residual LOD with CAM-derived equivalent LOD.

	\section{Methods}
	
	\subsection{Data and Observational Framework}

	LOD is the principal observable used here to characterize changes in axial rotation. The daily analysis uses the IERS C04 Earth Orientation Parameters (EOP) series from 1962 through 2024. These data are used for the daily LOD analysis and the UT1--UTC context. For multidecadal analysis, we use the annual values from the LUNAR97--IERS merged LOD product spanning 1883--2024. We use the published merged series directly; its construction and underlying historical observations are described in the cited sources. Its pre-space-geodetic segment is less precise than the modern IERS C04 record but provides the length needed to examine multidecadal variability \citep{bizouardCombinedSolutionC042009,stephensonLongtermFluctuations1995}.

	The CAM-derived equivalent LOD series available to us ends in 2022. To keep the sample interval identical for residual LOD and CAM, all reported harmonic fits, Lomb--Scargle spectra, wavelet diagnostics, BIC scans, LOD--CAM correlations, lag estimates, and post-1972 residual-slope metrics use the common interval 1883--2022. The 2023--2024 LOD values are retained only in the daily IERS C04 record and in figures showing the continuation of the observational LOD series.

	Residual LOD is formed by removing the IERS 2010 zonal-tide contribution, a linear secular tidal-braking term of $+2.3\,\mathrm{ms\,century^{-1}}$, and the available effective angular momentum (EAM) contribution. The arbitrary zero level of the secular correction is absorbed by the fitted intercept and therefore does not affect the estimated multidecadal period or amplitude. The common operational GFZ AAM, OAM, and HAM products used here span 1976--2024. Accordingly, the combined EAM correction is applied from 1976 onward and is unavailable for 1883--1975. Annual averaging attenuates the dominant short-period EAM variability, but unresolved low-frequency contributions may remain in the uncorrected historical interval. We therefore treat incomplete EAM coverage as a limitation rather than interpreting the residual as an exact separation of deep-interior processes.

	\subsection{Removal of Tidal Contributions}
	
	Solar and lunar tides deform the solid Earth and oceans and thereby produce periodic LOD variations. Long-period zonal tides, including the 18.6-year lunar nodal tide, together with shorter-period terms, can reach amplitudes of several milliseconds \citep{rayLunarSolarTorques1999}.
	
	We remove the zonal-tide contribution over the full daily interval using the IERS 2010 model for an elastic Earth with equilibrium oceans. These deterministic astronomical terms include the relevant lunar apsidal and nodal effects \citep{zotovEarthRotationVariations2026}.
	
	\subsection{Fluid Angular Momentum Contributions}

	Using Eq.~(1), we write the rotation vector as $\boldsymbol{\omega}=\Omega_0(m_1,m_2,1+m_3)$, where $m_1$ and $m_2$ describe polar motion, $p=m_1+i m_2$ is the corresponding complex polar-motion variable, and $m_3$ is the fractional change in axial rotation rate. To first order, the axial Liouville equation is
	\begin{equation*}
		C_m\Omega_0\dot{m}_3+\Omega_0\dot{\Delta I}_{33}+\dot{h}_3\simeq L_3,
	\end{equation*}
	where $C_m$ is the effective polar moment of inertia of the mantle and crust, $\Delta I_{33}$ is the axial inertia perturbation, $h_3$ is the axial relative angular momentum, and $L_3$ is the axial external torque. Thus, the differential equation contains $\dot{h}_3$. After removing the modeled external-torque contribution, integration of the anomaly equation relative to the reference mean state yields, with the integration constant absorbed into the mean rotation rate,
	\begin{equation}
		\frac{\Delta LOD(t)}{LOD_0}\simeq -m_3(t)\simeq \chi_3(t)
		=\frac{\Delta I_{33}(t)}{C_m}+\frac{h_3(t)}{C_m\Omega_0},
	\end{equation}
	where $LOD_0=86{,}400\,\mathrm{s}$ and $\Omega_0$ is Earth's mean angular velocity \citep{gross309EarthRotation2007,barnesAtmosphericAngularMomentum1983,petitIERSConventions2010}. The final relation contains $h_3$, rather than $\dot{h}_3$, because it describes the integrated angular-momentum anomaly. The expression is a first-order relation; the product-specific elastic and loading conventions are already included in the supplied excitation functions.

	We use the operational GFZ atmospheric (AAM v1.0), oceanic (OAM v1.0), and hydrological (HAM v1.2) excitation functions for 1976--2024 \citep{dobslawPredictingEarthOrientation2018,dobslawEAMProductDescription2019}. Each product provides dimensionless mass and motion terms. For each subsystem, the axial excitation is the sum of its mass and motion terms. The total EAM excitation, $\chi_3^{\mathrm{EAM}}$, is then obtained by summing the axial AAM, OAM, and HAM excitations and is converted to equivalent LOD through $\Delta LOD_{\mathrm{EAM}}=LOD_0\chi_3^{\mathrm{EAM}}$. The analysis therefore uses the provider-supplied excitation functions and does not reconstruct $\chi_3$ from raw angular-momentum fields.

	\subsection{Core Angular Momentum Inversion from Geomagnetic Fields}
	
	We infer angular-momentum variations in the fluid outer core within the frozen-flux approximation, which neglects magnetic diffusion and treats the magnetic field at the CMB as advected by the core flow \citep{jaultWestwardDriftCore1988,jacksonTimeDependentFlowCore1993a}. Under the geostrophic assumption, the inferred large-scale flow is tangential to the CMB and approximately aligned with contours of constant pressure \citep{asariMagneticEstimationEarths2015}.
	
	We use the COV-OBS.x2 geomagnetic-field model \citep{huderCOVOBSx2180Years2020}, which combines ground and satellite observations to estimate time-dependent Gauss coefficients for 1840--2020. CAM-derived equivalent LOD was calculated with WebGeodyn \citep{WebGeodyn}, maintained by the geodynamo group at Universit\'e Grenoble Alpes. The processed annual CAM-derived series available to us extends to 2022. Although the merged LUNAR97--IERS LOD record continues through 2024, all quantitative multidecadal fits and LOD--CAM comparisons reported in this study are restricted to the common 1883--2022 interval. The frozen-flux and geostrophic assumptions enter through this established inversion framework.
	
	Because the geodetic and CAM-derived series can differ in reference level, amplitude calibration, and inversion uncertainty, we compare their periods, phases, lagged correlations, and standardized low-frequency variations rather than requiring point-by-point agreement.

	\subsection{Harmonic Regression, Spectral Analysis, and Wavelet Diagnosis}

	All analyses in this subsection use annual data over the common 1883--2022 interval. We represent each multidecadal series with a linear trend plus a harmonic term,
	\begin{equation}
		LOD(t)=c_0+c_1(t-t_0)+A\cos\left(\frac{2\pi t}{P}\right)+B\sin\left(\frac{2\pi t}{P}\right)+\epsilon(t),
	\end{equation}
	where $P$ is the trial period, $c_0$ and $c_1$ are the intercept and linear coefficient, and $A$ and $B$ are the harmonic coefficients. The amplitude is
	\begin{equation}
		R=\sqrt{A^2+B^2},
	\end{equation}
	and the phase is calculated with the quadrant-aware function $\phi=\operatorname{atan2}(B,A)$. Trial periods from 50 to 100 yr are evaluated at 0.1-yr increments. For each trial period, the same four-parameter model is fitted by least squares and evaluated with
	\begin{equation*}
		\mathrm{BIC}=n\ln(\mathrm{RSS}/n)+k\ln n,
	\end{equation*}
	where $n$ is the number of annual observations, $\mathrm{RSS}$ is the residual sum of squares, and $k=4$ \citep{schwarzEstimatingDimensionModel1978}. Because $k$ is identical for all trial periods, the BIC penalty is constant across the scan; the criterion provides a consistent ranking of the candidate periods rather than a comparison of models with different dimensions.

	As an independent spectral diagnostic, we compute Lomb--Scargle periodograms of the linearly detrended annual series \citep{lombLeastSquaresFrequency1976,scargleStudiesAstronomicalTime1982}. Lomb--Scargle analysis is appropriate for a historical record assembled from different observing systems and remains valid under small departures from uniform sampling. A conventional FFT would instead require a uniformly sampled, gap-free record and would not supply a direct criterion for selecting among the candidate harmonic periods.

	We also use the continuous wavelet transform (CWT) to examine the time localization of multidecadal power. Short-period variability is reduced by removal of the modeled tidal component, subtraction of EAM where available, and annual averaging; no additional digital filter is applied before the harmonic, Lomb--Scargle, or wavelet analyses. For the Morlet wavelet,
	\begin{equation}
		\psi_0(\eta)=\pi^{-1/4}\exp(i\omega_0\eta)\exp(-\eta^2/2),
	\end{equation}
	with dimensionless central frequency $\omega_0=6$. Wavelet scale $s$ is converted to approximate Fourier period through
	\begin{equation}
		T=s\frac{4\pi}{\omega_0+\sqrt{2+\omega_0^2}}.
	\end{equation}
	The wavelet and coherence panels are used as qualitative time--frequency diagnostics; they are not treated as independent tests of a fixed phase relationship.

	\subsection{LOD--CAM Correspondence and UT1 Integration}

	Zero-lag and lagged correlations are calculated over 1883--2022. Positive lag denotes CAM leading LOD. Lagged correlations are evaluated from $-20$ to $+20$ yr at 1-yr increments. A centered 11-year moving average is shown only to compare the low-frequency shapes of the two series; it is not used to estimate the preferred period. Because both annual records are strongly autocorrelated, the raw, smoothed, and lagged correlations are interpreted as measures of correspondence rather than as independent tests of predictive significance. Smoothing further reduces the effective number of degrees of freedom \citep{chaoEstimatingCrossCorrelation2019}.

	The continuous UT1 contribution of the fitted LOD component is calculated from
	\begin{equation}
		\Delta UT1(t)\simeq-\int_{t_0}^{t}\frac{\Delta LOD(\tau)}{LOD_0}\,d\tau,
	\end{equation}
	where $\tau$ is expressed in seconds and $\Delta LOD$ is converted to seconds. Equivalently, a 1 ms LOD anomaly maintained for one year accumulates approximately $0.36525\,\mathrm{s}$ in UT1. This integration gives an order-of-magnitude contribution from the fitted multidecadal component; it does not simulate UTC step adjustments, operational bulletins, or short-term UT1 prediction errors.

	\section{Results and Discussion}
	
	\subsection{Post-1972 Reversal in Earth's Rotational Trend}
	
	The LOD record shows a marked change in trend near 1972. Over the analysis interval ending in 2022, the corresponding angular acceleration is $-13.1\times10^{-22}\,\mathrm{rad\,s^{-2}}$ before 1972 and $+50.6\times10^{-22}\,\mathrm{rad\,s^{-2}}$ thereafter. These values describe the broad trend in the original LOD record and are distinct from the residual-slope metric in Table~\ref{tab:lod_cam_metrics}, which is calculated after removal of the modeled contributions.
	
\begin{figure}[H]
		\centering
		\includegraphics[width=0.92\textwidth,height=0.82\textheight,keepaspectratio]{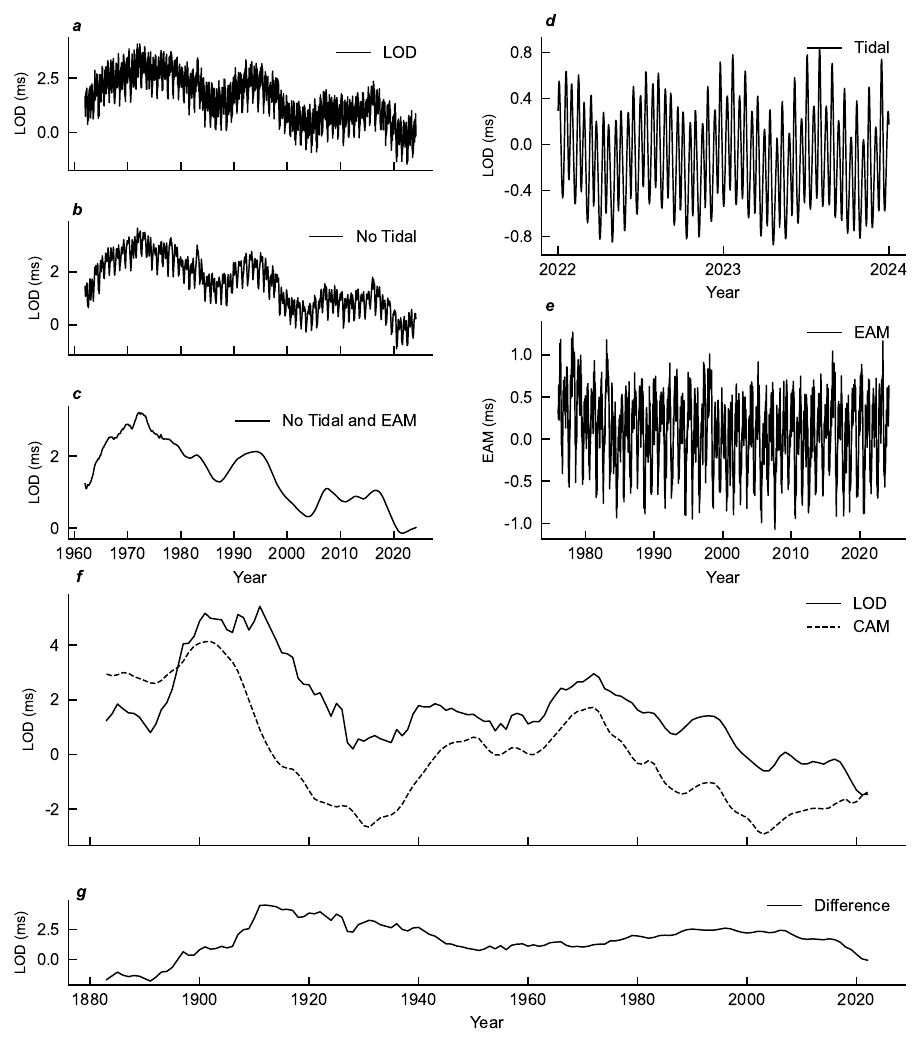}
		\caption{\textbf{Observed and modeled components of Earth's rotation variability.}
			\textbf{a}--\textbf{c}, Daily LOD anomalies in milliseconds (ms) derived from geodetic observations. \textbf{a}, Observed daily LOD variations. \textbf{b}, LOD after removing tidal effects. \textbf{c}, Residual LOD after subtracting the effective atmospheric, oceanic, and hydrological angular-momentum contributions and the long-term tidal-friction trend.
			\textbf{d}, Representative 2022--2024 segment of the modeled zonal-tide contribution, shown for visual clarity; the tidal correction was applied over the full daily interval.
			\textbf{e}, Combined EAM signal obtained by summing the axial AAM, OAM, and HAM excitation functions over their common 1976--2024 interval.
			\textbf{f}, Annual-mean comparison between the long-term residual LOD and the estimated CAM-derived equivalent LOD. The long-term residual is based on the 1883--2024 LUNAR97--IERS merged series; the LOD--CAM comparison is restricted to 1883--2022, and EAM corrections are limited to the intervals covered by the GFZ products.
			\textbf{g}, Difference between residual LOD and CAM-derived equivalent LOD.}
		\label{fig:Composite}
	\end{figure}

	The modeled lunisolar-tide and EAM contributions in Fig.~\ref{fig:Composite} are generally confined to the millisecond range and primarily affect periodic or short-term variability. In the adopted residual construction, these terms do not reproduce the observed low-frequency trend reversal.
	
	The timing and magnitude of the remaining trend are consistent with a low-frequency change in angular-momentum exchange involving the deep interior. Residual LOD broadly follows the low-frequency CAM-derived equivalent LOD in Fig.~\ref{fig:Composite}f, supporting a core-related contribution without identifying a unique coupling mechanism.

	\subsection{Relative Contributions to Long-Term LOD Trends}

	Table~\ref{tab:LOD_contributions} summarizes representative amplitudes and timescales of the main processes affecting LOD. The values are order-of-magnitude ranges compiled from the cited literature and standard Earth-rotation scaling; they are not independent estimates of each torque or excitation term. Tidal forcing and EAM dominate short-period and seasonal variability. Their principal components are oscillatory and strongly attenuated by annual averaging, although unresolved low-frequency EAM contributions cannot be excluded.
	
	Electromagnetic, gravitational, topographic, and viscous coupling at the CMB can transfer angular momentum on decadal to centennial timescales. The available observations do not permit a first-principles partition among these mechanisms. The empirical fraction reported below therefore measures only how much of the post-1972 residual LOD slope is represented by the fitted multidecadal component.
	
		Figure~\ref{fig:Composite} shows the successive removal of the tidal signal (panel b) and EAM (panel c). The resulting annual residual resembles the CAM-derived equivalent LOD (panel f). Their difference (panel g) includes observational error, incomplete surface-fluid corrections, and core-flow components not represented by the inversion.

	\subsection{An Approximately 70-Year Modulation and the Quantitative LOD--CAM Correspondence}

\begin{figure}[H]
		\centering
		\includegraphics[width=0.58\textwidth,height=0.78\textheight,keepaspectratio]{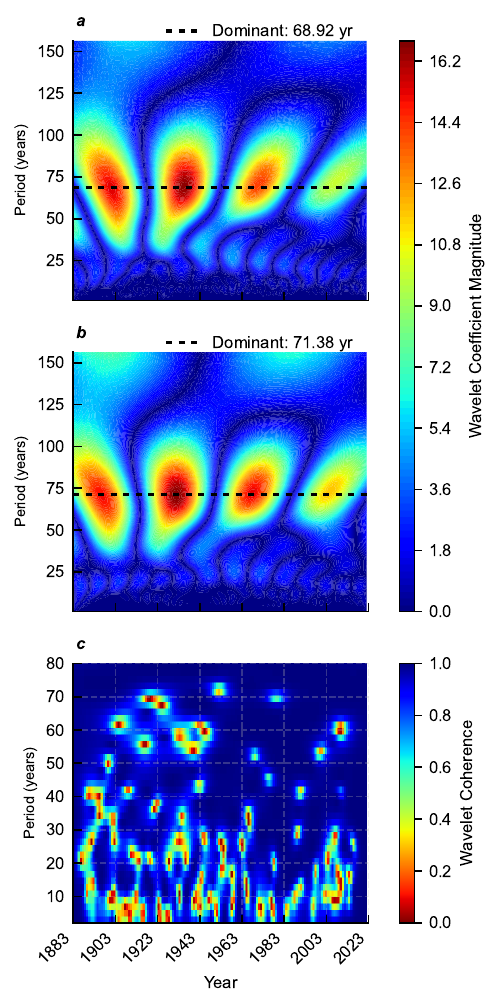}
		\caption{\textbf{Continuous wavelet transform and wavelet coherence of multidecadal LOD and CAM signals.}
			\textbf{a}, Morlet wavelet transform of annually averaged residual LOD after removal of modeled tidal and available EAM components, showing a dominant period near 68.9 years.
			\textbf{b}, Wavelet spectrum of CAM-derived equivalent LOD, revealing a similar peak near 71.4 years.
			\textbf{c}, Wavelet coherence between LOD and CAM. Intermittent regions of elevated coherence occur in the 60--75 yr band; the panel is interpreted as a qualitative time--frequency diagnostic and not as an independent test of a stable phase relationship.}
		\label{fig:Wavelet}
	\end{figure}

	The wavelet transform of annually averaged residual LOD shows maximum multidecadal power near 69 yr (Fig.~\ref{fig:Wavelet}a). Harmonic regression and Lomb--Scargle analysis identify a similar period. CAM-derived equivalent LOD also contains a multidecadal component, with maximum power near 71 yr (Fig.~\ref{fig:Wavelet}b).
	
	Elevated wavelet coherence occurs intermittently in the 60--75 yr band, particularly during parts of the twentieth century (Fig.~\ref{fig:Wavelet}c). This panel supports common multidecadal variability but, without a separate phase-significance analysis, is not used to infer a stable phase relationship.
	
	This correspondence is consistent with angular-momentum exchange across the CMB. Possible contributors include torsional oscillations and other large-scale magnetohydrodynamic motions in the fluid core \citep{jaultWestwardDriftCore1988,braginskyTorsionalMagnetohydrodynamic1970,gilletFastTorsionalWaves2010,moundMechanismsCoreMantle2005}. The interpretation is not unique, and the analysis does not identify a single coupling mechanism. Seismic studies have also reported an approximately 70-year modulation of inner-core rotation from body-wave travel-time changes \citep{yangMultidecadalVariationEarth2023,wangInnerCoreBacktracking2024a}, showing that multidecadal variability has been identified in another deep-Earth observable.

\begin{figure}[H]
		\centering
		\includegraphics[width=0.72\textwidth,height=0.78\textheight,keepaspectratio]{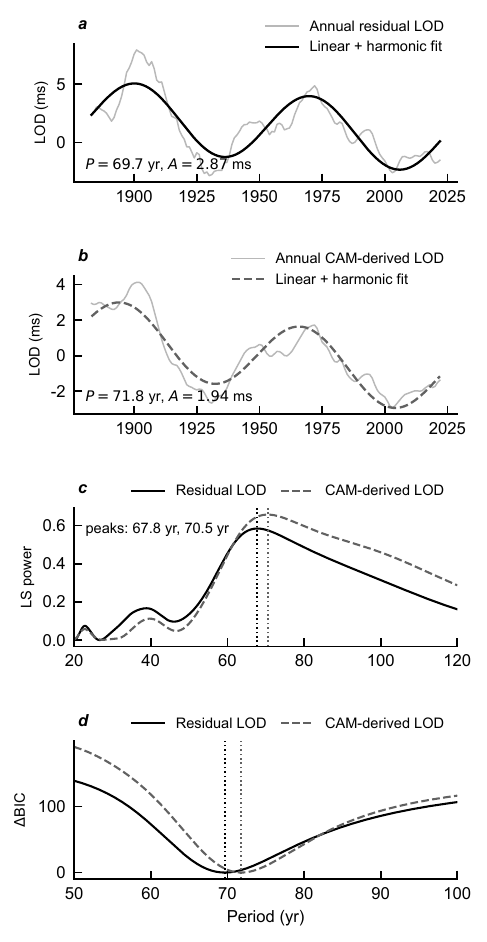}
		\caption{\textbf{Quantitative identification of the multidecadal component in residual LOD and CAM-derived equivalent LOD.}
			\textbf{a}, Residual LOD and the best-fitting harmonic model over 1883--2022, obtained from the 50--100 yr period scan.
			\textbf{b}, CAM-derived equivalent LOD and the corresponding harmonic fit over 1883--2022.
			\textbf{c}, Lomb--Scargle spectra of linearly detrended residual LOD and CAM-derived equivalent LOD.
			\textbf{d}, BIC-based period scan. Both harmonic regression and Lomb--Scargle spectra indicate preferred periods close to 70 yr, although the limited number of cycles requires interpretation as an approximate multidecadal modulation rather than a precisely fixed periodicity.}
		\label{fig:PeriodQuant}
	\end{figure}

	The fitted parameters, all obtained over 1883--2022, are summarized in Fig.~\ref{fig:PeriodQuant} and Tables~\ref{tab:quantitative_components} and \ref{tab:lod_cam_metrics}. Harmonic regression gives preferred periods of 69.7 yr for residual LOD and 71.8 yr for CAM-derived equivalent LOD, with amplitudes of 2.87 and 1.94 ms and maxima near 1970.45 and 1967.01, respectively. The corresponding Lomb--Scargle peaks are 67.8 and 70.5 yr, and the BIC scans favor periods near 70 yr rather than 60 or 90 yr. Because the records contain only about two complete multidecadal cycles, these estimates are best interpreted as a 60--80 yr modulation with a preferred timescale near 70 yr, not as a stationary oscillation with a uniquely determined period.
	
	Using the best-fitting linear-plus-harmonic model with $t_0=1952.5$ yr, the curves shown in Fig.~\ref{fig:PeriodQuant} can be written explicitly as
	\begin{align}
		LOD_{\mathrm{res}}(t) &= 1.385 - 0.0156(t-1952.5) - 0.367\cos\left(\frac{2\pi t}{69.7}\right) + 2.850\sin\left(\frac{2\pi t}{69.7}\right),\\
		LOD_{\mathrm{CAM}}(t) &= -0.034 - 0.0189(t-1952.5) - 1.537\cos\left(\frac{2\pi t}{71.8}\right) + 1.181\sin\left(\frac{2\pi t}{71.8}\right),
	\end{align}
	where LOD is in milliseconds and $t$ is decimal year. The harmonic terms can equivalently be written as $2.874\cos[2\pi(t-1970.45)/69.7]$ and $1.939\cos[2\pi(t-1967.01)/71.8]$, making the fitted maxima explicit.
	
\begin{figure}[H]
		\centering
		\includegraphics[width=0.72\textwidth,height=0.72\textheight,keepaspectratio]{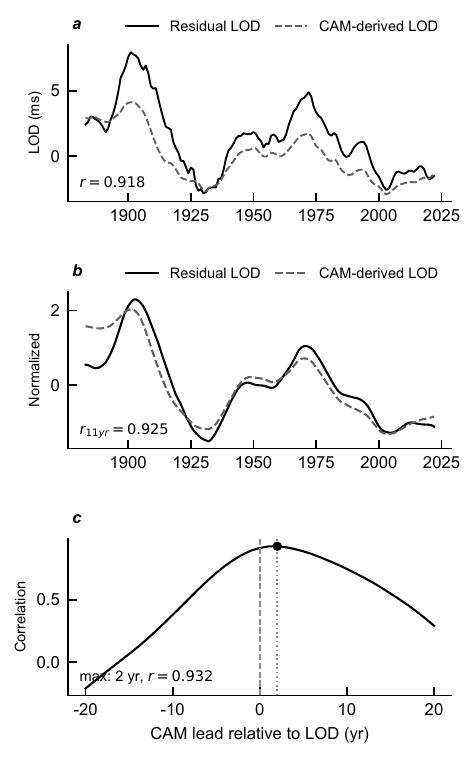}
		\caption{\textbf{Multidecadal correspondence between residual LOD and CAM-derived equivalent LOD.}
			\textbf{a}, Annual residual LOD and CAM-derived LOD over 1883--2022.
			\textbf{b}, Standardized components after applying a centered 11-year moving average, shown only as a descriptive comparison of low-frequency behavior.
			\textbf{c}, Lagged correlation as a function of CAM lead relative to LOD, evaluated from $-20$ to $+20$ yr in 1-yr increments. The correlation maximum is broad for CAM leads of approximately 1--3 yr, with the numerical maximum at 2 yr.}
		\label{fig:LODCAMRelation}
	\end{figure}

	Over the common interval 1883--2022, annual residual LOD and CAM-derived LOD have a zero-lag correlation of 0.918 (Fig.~\ref{fig:LODCAMRelation}). Their standardized, centered 11-year moving averages have a correlation of 0.925, reported only as a descriptive low-frequency measure because smoothing reduces the effective degrees of freedom \citep{chaoEstimatingCrossCorrelation2019}. The lagged-correlation curve has a broad maximum for a CAM lead of approximately 1--3 yr, with a numerical maximum of 0.932 at 2 yr. Given the strong autocorrelation of both series, these values describe the alignment of the low-frequency variability in the two series and are not used to assess predictive significance or to infer a unique causal mechanism.

	\subsection{Implications for Earth System Coupling and Future Leap Seconds}
	
	Residual LOD and CAM-derived equivalent LOD contain similar multidecadal variability, supporting a core-related contribution to low-frequency variability in Earth's rotation. The analysis does not determine how that contribution is partitioned among electromagnetic, topographic, gravitational, and viscous coupling at the CMB.
	
\begin{figure}[H]
		\centering
		\includegraphics[width=0.68\textwidth,height=0.62\textheight,keepaspectratio]{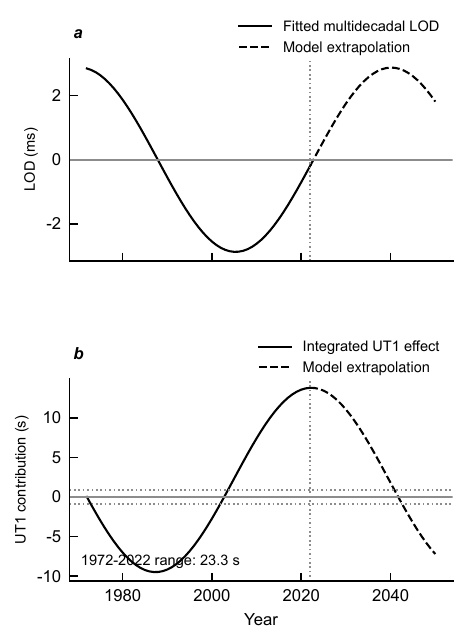}
		\caption{\textbf{Order-of-magnitude contribution of the fitted multidecadal LOD component to continuous UT1.}
			\textbf{a}, Fitted multidecadal residual LOD component and its model extrapolation beyond the observed interval.
			\textbf{b}, Continuous UT1 contribution obtained by integrating the fitted LOD component. The horizontal dashed lines mark the $\pm$0.9 s scale relevant to the usual UT1--UTC tolerance. This calculation is not an operational leap-second forecast and does not simulate UTC step adjustments.}
		\label{fig:UT1Relevance}
	\end{figure}

	Because UT1 is the time integral of LOD anomalies, the fitted multidecadal component produces a continuous UT1 contribution (Fig.~\ref{fig:UT1Relevance}). Integration from 1972 gives a peak-to-peak range of approximately 23.3 s over 1972--2022. This magnitude shows that the fitted component is relevant to UT1--UTC variability, although it represents only one contribution to the observed signal.
	
	This integration is not a prediction of leap-second insertions. UTC is adjusted in discrete steps, whereas Fig.~\ref{fig:UT1Relevance} shows only the continuous UT1 contribution of the fitted multidecadal component. Operational decisions also depend on short-term UT1 forecasts, other LOD variations, and the applicable timekeeping policy.
	
	In the fitted empirical model, the multidecadal component does not indicate sustained near-term shortening of the day that would, by itself, require a negative leap second. Extrapolation of the fitted residual-LOD harmonic beyond the observed interval reaches its next positive maximum near 2040. The extrapolated multidecadal component alone therefore does not support an imminent negative leap second. This conclusion is conditional on the fitted model and is not an operational UTC forecast; it is consistent with recent projections that also find no imminent need for a negative leap second \citep{malkinShouldExpectFurther2024}.
	
\begin{figure}[H]
		\centering
		\safeincludegraphics[width=0.75\textwidth,height=0.76\textheight,keepaspectratio]{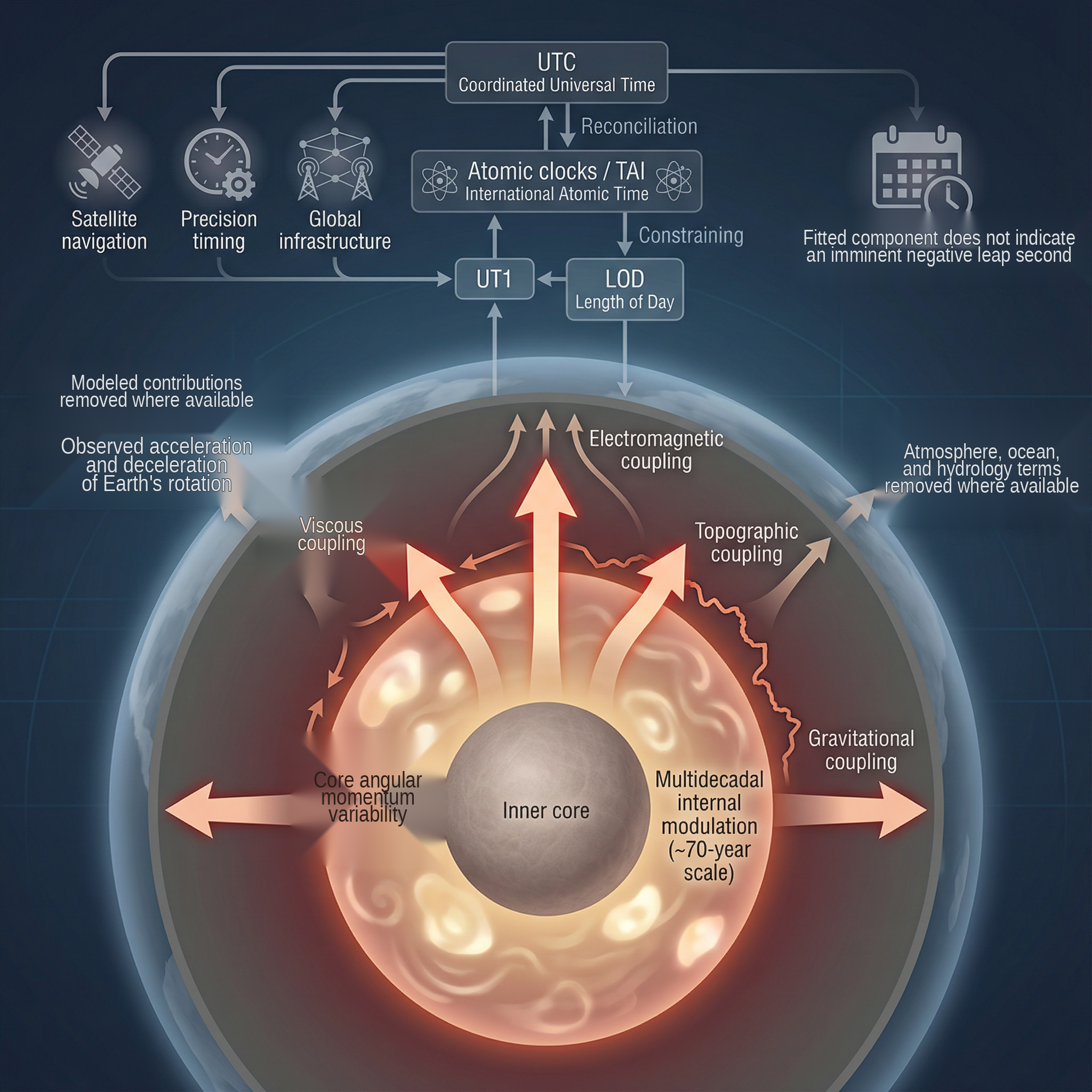}
		\caption{\textbf{Conceptual framework linking deep-Earth variability, Earth rotation, and timekeeping.} Multidecadal core-related angular-momentum variability may contribute to residual LOD changes through core--mantle coupling. After removing modeled tidal terms and available atmosphere--ocean--hydrology angular-momentum contributions, the residual LOD contains a multidecadal component that is consistent with CAM-derived equivalent LOD. Through time integration, this LOD component contributes to UT1 and is therefore relevant to leap-second discussions. The diagram is conceptual and does not imply a unique partition among electromagnetic, topographic, gravitational, and viscous CMB coupling.}
		\label{fig:Conceptual}
	\end{figure}

	Figure~\ref{fig:Conceptual} summarizes this interpretation. After removal of the modeled tides and available atmosphere--ocean--hydrology contributions, residual LOD retains a multidecadal component consistent with CAM-derived equivalent LOD. This correspondence relates deep-interior angular-momentum variability to UT1 while leaving the partition among CMB coupling mechanisms unresolved.
	
	The interpretation has several limitations. The LUNAR97--IERS series combines historical and modern observations, and the pre-space-geodetic segment is less precise than the modern EOP record. The GFZ EAM products used here begin in 1976, so no EAM correction is available for the 1883--1975 portion of the LOD record. Annual averaging attenuates the dominant short-period EAM variability, but uncorrected low-frequency contributions may remain in that historical interval. The record resolves only about two complete multidecadal cycles, so the preferred period near 70 yr is approximate. CAM-derived LOD also depends on the assumptions and regularization of the geomagnetic inversion. Finally, the UT1 integration quantifies the contribution of a single fitted LOD component rather than the complete observed UT1 signal and is not an operational forecast of UTC adjustments.
	
	The results support a core-related contribution to long-term variability in Earth's rotation and place the associated UT1 changes in a geophysical context (Fig.~\ref{fig:Conceptual}). Any extrapolation beyond the observed interval remains conditional on the persistence of the inferred multidecadal structure and on the adopted residual construction.

		\section*{Declarations}
	
	\subsection*{Availability of data and materials}
	The daily LOD and UT1 series are from the IERS C04 EOP solution for 1962--2024. Tidal contributions were removed with the IERS 2010 model where applicable. The long-term input is the annual LUNAR97--IERS merged LOD series for 1883--2024, while the reported multidecadal fits and LOD--CAM statistics use 1883--2022. Operational GFZ AAM v1.0, OAM v1.0, and HAM v1.2 excitation functions are used over their common 1976--2024 interval. CAM-derived equivalent LOD was obtained from the COV-OBS.x2 geomagnetic-field model through WebGeodyn, maintained by the geodynamo group at Universit\'e Grenoble Alpes.
	
	The primary data are publicly available from IERS, GFZ, and the COV-OBS.x2 and WebGeodyn providers \citep{WebGeodyn,IERS_C04}. The processed series and analysis scripts are available from the corresponding author upon reasonable request.
	
	\subsection*{Competing interests}
	The authors declare that they have no competing interests.
	
	\subsection*{Funding}
	The authors gratefully acknowledge financial support from the Strategic Priority Research Program of the Chinese Academy of Sciences (Grant No. XDB1070202).
	
	\subsection*{Authors' contributions}
	ZZ: Data curation, formal analysis, visualization, and writing--original draft.  
	YW: Conceptualization, supervision, and writing--review and editing.  
	XL: Data validation and analysis of the EOP and geomagnetic-inversion products.  
	DY: Conceptual discussion and writing--review and editing.  
	XC: Conceptual discussion and review of the manuscript structure.  
	XY: Supervision and writing--review and editing.  
	SZ: Supervision, interpretation of the results, and writing--review and editing.
	
	\subsection*{Acknowledgements}
	The authors thank Benjamin Fong Chao (Shandong University of Science and Technology), Pengshuo Duan (Shanghai Astronomical Observatory, Chinese Academy of Sciences), and Yi Yang (Nanjing University) for discussions and comments on earlier versions of the manuscript.
	\clearpage
\section*{Tables}
	
	\begin{table}[H]
        \centering
        \caption{Representative roles of major geophysical processes in long-term LOD variability.}
        \label{tab:LOD_contributions}
        \renewcommand{\arraystretch}{1.12}
        \footnotesize
        \begin{tabular}{@{}L{3.0cm} L{2.3cm} L{3.0cm} L{5.0cm}@{}}
            \toprule
            \textbf{Process} & \textbf{Typical LOD scale or trend} & \textbf{Characteristic timescale} & \textbf{Role in long-term residual LOD} \\
            \midrule
            Tidal effects & $\sim10^{-3}$ s & Sub-daily to interannual & Large modeled periodic component; removed before residual analysis \\
            Oblateness variation ($J_2$) & $\sim10^{-6}$ s & Decadal to centennial & Minor contribution to recent low-frequency residual trend \\
            Effective angular momentum & $\sim10^{-3}$ s & Daily to decadal & Dominant short-term and seasonal component; limited by data coverage \\
            Long-term tidal braking & $+2.3$ ms century$^{-1}$ & Centennial to millennial & Secular background removed in residual construction \\
            Core angular momentum exchange & $10^{-3}$--$10^{-2}$ s & Decadal to centennial & Plausible source of multidecadal residual variability; exact coupling partition not quantified \\
            \bottomrule
        \end{tabular}
        \par\vspace{0.5em}
        \begin{minipage}{0.98\textwidth}
        \footnotesize
        \textit{Note.} The listed scales are representative order-of-magnitude ranges compiled from the cited literature and standard Earth-rotation scaling. They provide context for the residual construction and do not constitute a first-principles partition among individual CMB coupling mechanisms. Representative sources are: tidal effects \citep{gross309EarthRotation2007,rayLunarSolarTorques1999,petitIERSConventions2010}; oblateness variation \citep{zotovAnalysisEarthPolar2022,mitrovicaReconcilingChangesEarth2015,yoderSecularVariationEarth1983a}; effective angular momentum \citep{gross309EarthRotation2007,yanEffectGlobalMass2012,dobslawPredictingEarthOrientation2018}; long-term tidal braking \citep{stephensonLongtermFluctuations1995,rayLunarSolarTorques1999,stephensonMeasurementEarthRotation2016}; and core angular momentum exchange \citep{buffettGeomagneticFluctuationsReveal2014,duanIntradecadalVariationsLength2020,jaultWestwardDriftCore1988,gilletFastTorsionalWaves2010}.
        \end{minipage}
    \end{table}
	
	\begin{table}[H]
		\centering
		\caption{Quantitative properties of the fitted multidecadal components in residual LOD and CAM-derived equivalent LOD.}
		\label{tab:quantitative_components}
		\renewcommand{\arraystretch}{1.15}
		\footnotesize
		\begin{tabular}{@{}L{3.1cm} L{2.2cm} L{2.0cm} L{1.8cm} L{2.4cm}@{}}
			\toprule
			\textbf{Series} & \textbf{Harmonic period (yr)} & \textbf{Amplitude (ms)} & \textbf{Peak year} & \textbf{Lomb--Scargle peak (yr)} \\
			\midrule
			Residual LOD & 69.7 & 2.87 & 1970.45 & 67.8 \\
			CAM-derived LOD & 71.8 & 1.94 & 1967.01 & 70.5 \\
			\bottomrule
		\end{tabular}
		\par\vspace{0.5em}
		\begin{minipage}{0.98\textwidth}
		\footnotesize
		\textit{Note.} The values are empirical estimates from harmonic regression over the 50--100 yr period range and from Lomb--Scargle spectra of linearly detrended annual series. They should not be interpreted as a partition among specific CMB coupling mechanisms.
		\end{minipage}
	\end{table}
	
	\begin{table}[H]
		\centering
		\caption{LOD--CAM correspondence and post-1972 residual LOD trend metrics.}
		\label{tab:lod_cam_metrics}
		\renewcommand{\arraystretch}{1.12}
		\begin{tabular}{@{}L{8.5cm}L{5cm}@{}}
			\toprule
			\textbf{Quantity} & \textbf{Value} \\
			\midrule
			Zero-lag correlation between residual LOD and CAM-derived LOD & 0.918 \\
			Centered 11-year moving-average correlation & 0.925 \\
			Maximum lagged correlation & 0.932 \\
			Lag of broad correlation maximum & CAM leads LOD by approximately 1--3 yr; numerical maximum at 2 yr \\
			Post-1972 residual LOD slope & $-0.108$ ms yr$^{-1}$ \\
			Slope of fitted multidecadal component over same interval & $-0.099$ ms yr$^{-1}$ \\
			Empirical fraction of residual slope represented by fitted component & approximately 92\% \\
			Angular acceleration equivalent of post-1972 residual slope & $28.9\times10^{-22}$ rad s$^{-2}$ \\
			\bottomrule
		\end{tabular}
		\par\vspace{0.5em}
		\begin{minipage}{0.98\textwidth}
		\footnotesize
		\textit{Note.} The post-1972 fraction is the ratio of two fitted slopes over the same interval. It is descriptive, sensitive to the residual construction and selected interval, and is not a physical decomposition of electromagnetic, topographic, gravitational, or viscous coupling. The centered 11-year moving-average correlation is included only as a descriptive low-frequency diagnostic and is not used as an independent significance test.
		\end{minipage}
	\end{table}
	
	\clearpage
\section*{Figures}
	
\end{document}